# Integrated GHz silicon photonic interconnect with micrometer-scale modulators and detectors

**Long Chen, Kyle Preston, Sasikanth Manipatruni, and Michal Lipson**

*School of Electrical and Computer Engineering, Cornell University, Ithaca, NY 14853*

*lc286@cornell.edu*

**Abstract:** We report an optical link on silicon using micrometer-scale ring-resonator enhanced silicon modulators and waveguide-integrated germanium photodetectors. We show 3 Gbps operation of the link with 0.5 V modulator voltage swing and 1.0 V detector bias. The total energy consumption for such a link is estimated to be ~120 fJ/bit. Such compact and low power monolithic link is an essential step towards large-scale on-chip optical interconnects for future microprocessors.

**OCIS codes:** (200.4650) Optical interconnects; (250.5300) Photonic integrated circuits; (130.4110) Modulators; (040.5160) Photodetectors.

Silicon photonic circuits are considered a promising solution to provide the large communication bandwidth and low power consumption required for on-chip interconnects in future microprocessors [1-4]. A tremendous amount of progress has been reported on the development of discrete photonic components on silicon, including light sources [5-7], switching and routing networks [8,9], electro-optical modulators based on silicon, SiGe, or III-V compounds [10-16], and photodetectors using III-V compounds, implanted silicon, or germanium [17-24]. Large scale integration of these various photonic components within the footprint and power constraints of on-chip interconnects is of paramount importance [25]. In particular, the integration of compact and low power modulators and photodetectors – the two key interfaces bridging the electrical and optical domains – on the same chip is crucial for a fully functional optical interconnect system [1-4]. Similar integration has been previously reported, however only with millimeter-scale modulators and relatively large power consumption [26]. Here we report an on-chip optical link using a silicon modulator and a germanium photodetector which are waveguide-integrated, monolithically fabricated and micrometer-scale. We demonstrate 3 Gbps data communication with 0.5 V modulator voltage swing and 1.0 V detector bias, and a total energy consumption of only ~120 fJ/bit. The basic interconnect link presented here can serve as a platform for developing scalable high-bandwidth and low-power optical interconnect architectures.

We designed a proof-of-concept on-chip optical link with point-to-point optical data communication. The scheme is illustrated in Figure 1 with scanning electron microscope (SEM) images of the actual device. As shown in Figure 1(a), a silicon waveguide connects the optical output of a silicon electro-optic modulator (see Figure 1(b)) to the optical input of a waveguide-integrated germanium detector (see Figure 1(c)). In an interconnect system the electrical input of the modulator and the electrical output of the detector of such a link would be both connected to microelectronic circuits.

To reduce the size and power consumption of the link, we choose a resonator-based modulator of only 12 μm in diameter. The modulator consists of a bus waveguide and a microring resonator with a lateral PIN junction across the waveguide forming the ring [11], as shown in Figure 1(b). The resonator scheme greatly enhances the sensitivity of the optical signal to the small index change in the ring induced by the injection and extraction of free carriers. This enhanced sensitivity, together with the small size of the ring, leads to much lower power consumption than non-resonator-based modulators [10,12,25]. We have previously demonstrated up to 18 Gbps operation using such modulators [13]. For the detector (see Figure 1(c)) we choose a waveguide-integrated metal-germanium-metal detector due to its ultra-low capacitance and fast response. We have recently demonstrated detectors with 2.4 fF capacitance and response time as short as 8.8 ps using a similar design [24].

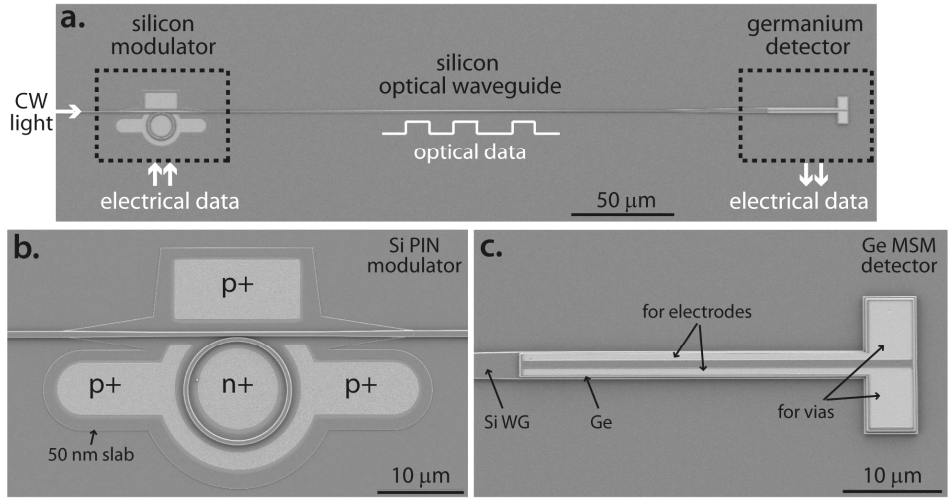

Fig. 1. Integrated optical interconnect link with a silicon electro-optical modulator, silicon waveguide, and germanium-on-silicon photodetector, as illustrated with SEM images of (a) the full optical link, (b) the ring resonator modulator after silicidation, and (c) the lateral metal-germanium-metal photodetector after electrode hole etch.

The entire chip is fabricated using wafer bonding and standard monolithic processes. Starting from a 4-inch silicon-on-insulator (SOI) wafer and a 4-inch crystalline germanium wafer, we use an ion-assisted layer cutting technique [27] to transfer a thin germanium film from the bulk wafer onto the SOI with a thin SiO$_2$ buffer layer. The schematic is shown in Figure 2(a), with the inset showing an SEM of the obtained GeOI-SOI wafer (germanium-on-insulator on silicon-on-insulator). The details can be found in ref. [21,24] and references therein. The silicon layer is 260 nm thick (used for the passive waveguides and the microring modulator), and the germanium layer is 260 nm thick (used only in the detector region). The SiO$_2$ layer between silicon and germanium is about 80 nm thick for electrical isolation and control of optical coupling. After the layer transfer, the fabrication consists of multiple steps of electron-beam lithography, etching, implantation and deposition, and it is illustrated in Figure 2(b). After etching alignment marks, we pattern the germanium for the detector and remove the germanium everywhere else (step (1)). We then deposit 120 nm of SiO$_2$ to protect the germanium from subsequent processing. Next we define the silicon waveguides and the ring resonator and partially etch the silicon layer, leaving a 50 nm silicon slab (step (2)). Another lithography step is used to cover the modulator region and continue the silicon etch, leaving the 50 nm slab only around the modulator to form the PIN diodes (step (3)). We then deposit 20 nm of SiO$_2$ for silicon passivation, and pattern the electrical contacts for the modulator. We perform the implantation steps for the p-region (outside the ring, BF$_2$, dose 3E15 at 45 KeV) and n-region (inside the ring, Phosphorous, dose 2E15 at 33 KeV) respectively. Due to the presence of germanium a relatively low temperature anneal (650$^\circ$C rapid thermal anneal for 2 mins) is used to activate the dopants. About 15 nm of nickel is then deposited in the doped region and annealed at 550$^\circ$C for 50 second to form the nickel-silicide (step (4)). The modulator after this step is shown in Figure 1(b), where the slab surrounding the modulator and the silicide regions are clearly visible. After this, we pattern electrode hole openings for the detector (see Figure 1(c)), and evaporate 10 nm titanium followed by 400 nm aluminum to form metal electrodes for the detector and also on top of the modulator silicide (step (5)). We then deposit 1 μm SiO$_2$ top cladding, pattern via holes and contact pads connecting to the aluminum electrodes of the modulator and detector (step (6)). The sample is then diced and facet-polished for optical and electrical testing. The waveguide forming the ring resonator has a width of 450 nm and a coupling gap of 200 nm to the bus waveguide. The detector has a length of approximately 40 μm, and the gap between the two planar electrodes is 700 nm. We estimate the capacitance of the detector junction (without the contact pads) to be only 4 fF, which would lead to low electrical power receivers by allowing for a large load resistance and relaxing the gain requirement on the electrical amplifiers [28,29]. The footprints of the modulator and the photodetector are about 115 μm$^2$ and 125 μm$^2$, respectively. The waveguide linking the modulator and photodetector is about 250 μm in length.

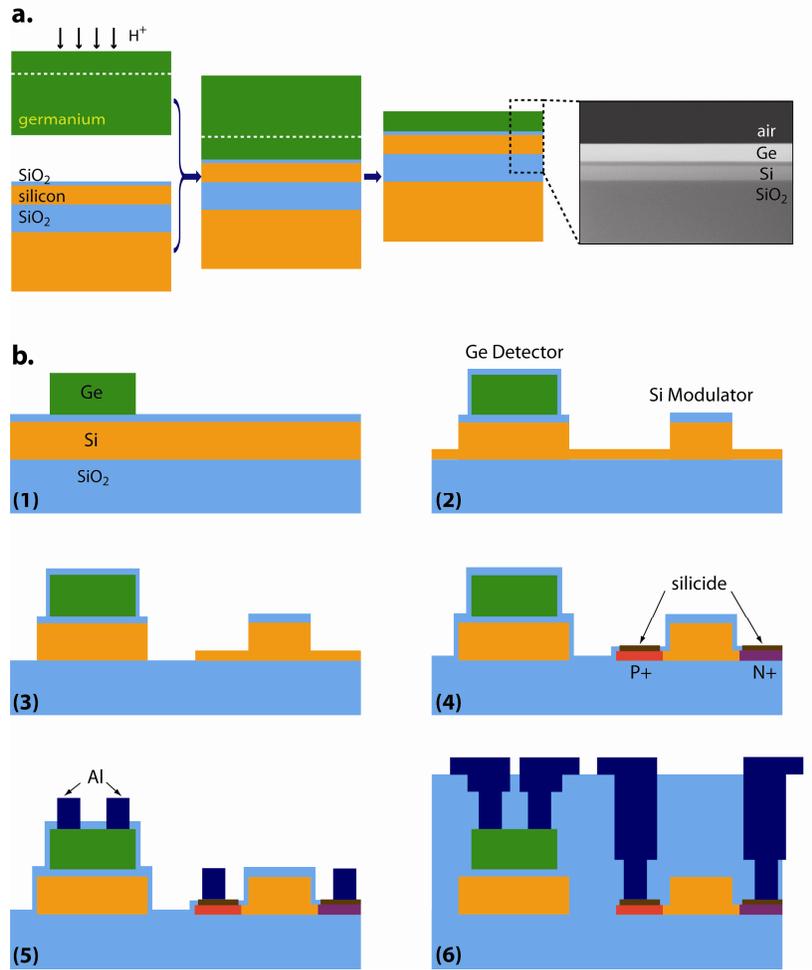

Fig. 2. Fabrication processes of the integrated optical link. (a) Wafer bonding and ion-assisted layer cutting technique to transfer crystalline germanium onto SOI to obtain the GeOI-SOI wafer. (b) Standard monolithic lithography steps to pattern the silicon modulator and the germanium detector on the same chip.

We characterize the DC behavior of the PIN silicon modulator and MSM germanium detector and show low DC resistance for the modulator and low dark current for the detector. A diagram of the experimental setup is shown in Figure 3(a). The optical path is shown with thicker red line and consists of a continuous-wave laser, fiber polarization controller (PC), and the silicon waveguide on the chip from the facet to the modulator and between the modulator and the detector. The electrical path is shown with thinner black line. Note that while external electronics are used here, in principle the receiver and the driver functions can be monolithically integrated on the same silicon chip [26]. Figure 3(b) shows the IV response of the modulator PIN diode with differential resistance of approximately 2 kΩ at 1.6 V and 770 Ω at 2 V. Figure 3(c) shows an example of the detector photocurrent and dark current as a function of detector bias voltage. The dark current is 280 nA at 0.5 V bias and 1.4 μA at 1 V. When the waveguide is illuminated with 0.67 mW optical power (transverse-electrical polarized, measured from the fiber tip) at λ = 1530 nm, the photocurrent saturates at around 0.13 mA with a bias voltage of 0.3 V. The measured responsivity of approximately 0.2 A/W is limited by optical losses throughout the propagation path. From similar devices we estimate 4 dB coupling loss (without nanotaper spot-size converter), 1 dB insertion loss through the modulator (including the two transitions between the regions with and without 50 nm slab), and 2 dB total waveguide propagation loss before the detector, resulting in an estimated responsivity of around 0.8 – 0.9 A/W.

We also measure the optical transmission spectrum of the ring resonator using the germanium detector at the waveguide termination. The detector is biased here at $V_{det}$ = 0.5 V. The spectrum is shown in Figure 2(d), showing clear resonances of the ring modulator. The resonances are slightly under-coupled and exhibit moderate extinction ratios from 7 to 9 dB. The resonance quality factors Q (defined as the ratio of the resonance wavelength $\lambda_r$ to the linewidth $\lambda_{FWHM}$) are around 5,000 - 6,500. Note that while the response falls off at longer wavelength due to the decreasing absorption of germanium beyond its direct band gap, throughout the C-band (1530 - 1565 nm) the photocurrent is about two orders of magnitude greater than the dark current, providing a wide operating wavelength range. Further improvement can be achieved by doping the germanium and shifting the metal electrodes away from the optical mode to avoid the competing metal absorption.

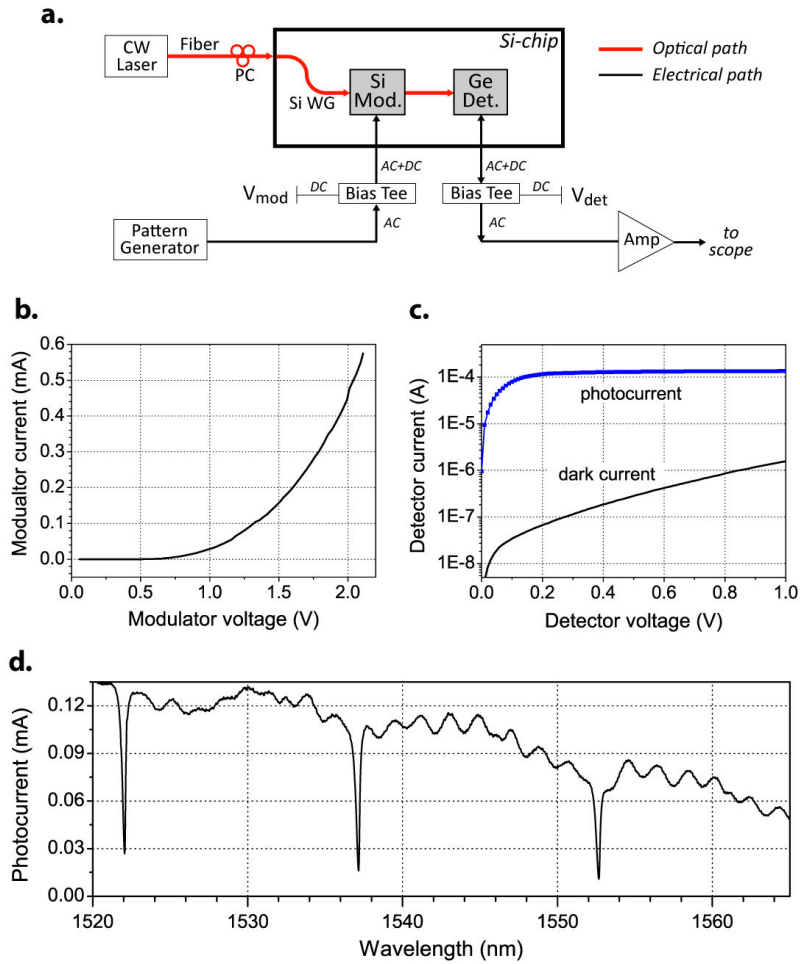

Fig. 3. Characterization of the integrated optical link. (a) Experimental setup illustrating the optical and electrical paths. (b) IV curve of silicon electro-optical modulator. (c) Dark and illuminated (0.67 mW from fiber) IV curve of germanium photodetector. (d) DC spectrum of detector response with 0.67 mW fiber illumination and 0.5 V detector bias, showing the resonances of the ring modulator before the detector.

With this integrated optical link we demonstrate open eye diagrams at 3 Gbps data transmission. The modulation is controlled by applying a forward bias voltage to the PIN diode which injects free carriers into the resonator. This shifts the resonant wavelength due to the free-carrier dispersion [30], and therefore changes the transmission of the optical signal at resonance. In our measurements the CW optical input is tuned to one of the ring resonances at λ ~ 1522 nm, which has a linewidth of ~0.23 nm, a quality factor of ~6,500, and an extinction ratio of ~7 dB. A non-return-to-zero (NRZ) electrical signal from a $2^{31}$-1 pseudorandom pattern generator is applied to the modulator PIN with a peak-to-peak voltage swing of 0.5 $V_{pp}$ and DC bias of 1.4 V (see Figure 3(a)). This signal drives the modulator diode on and off, thus imprinting the electrical data onto the optical signal. After transmitting through the silicon waveguide, the optical signal is then converted back to an electrical signal at the detector block, where the photocurrent is recorded with a sampling oscilloscope after a detector bias circuit and a low noise amplifier. In Figure 3(a) and (b) we compare the electrical input applied to the modulator and the electrical output retrieved from the photodetector, at a data rate of 3 Gbps. Figure 3(c) shows an open electrical eye diagram of the received data at the detector block. The optical power coupled into the waveguide is estimated to be approximately 0.6 mW, and the averaged output DC current of the detector is about 0.1 mA.

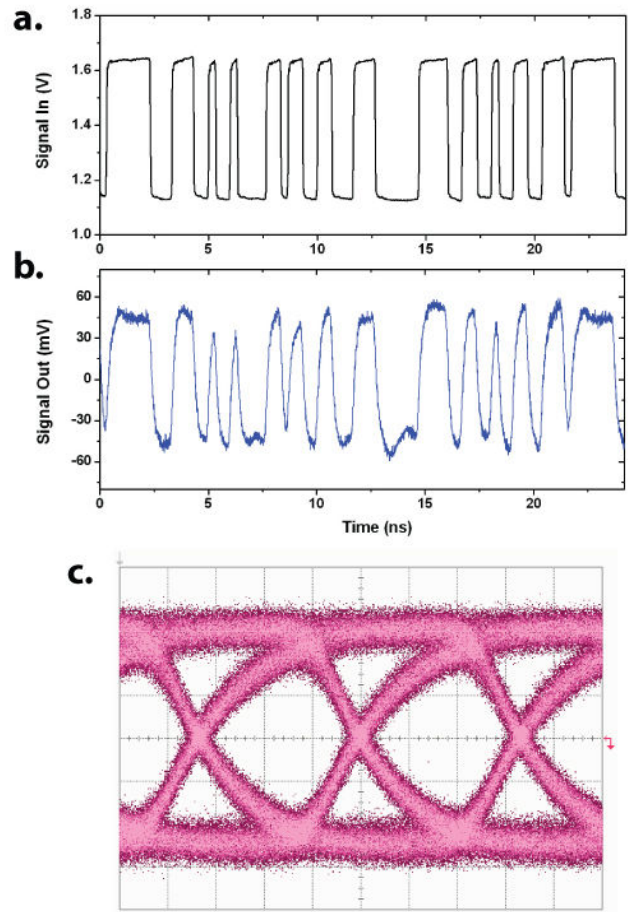

Fig. 4. AC characterization of the interconnect link with 1.4 V DC bias and 0.5 $V_{pp}$ signal driving the modulator, a detector bias of 1.0 V, and averaged detector output DC current of ~0.1 mA. (a) Input electrical signal applied to the modulator. (b) Output electrical signal amplified from the photodetector. (c) Eye diagram of the output electrical signal for $2^{31}$-1 NRZ PRBS at 3 Gbps.

The integrated optical link demonstrated here with micrometer-scale modulators and detectors operates at low power of approximately 120 fJ/bit. We estimate the power consumption of the compact modulator, including both the AC switching energy and the DC holding energy, to be approximately 86 fJ/bit at 3 Gbps data rate. The switching energy is estimated based on the optical resonance shift and the carrier injection as follows. Assuming a resonance shift ($\delta\lambda_r$) equal to the optical linewidth ($\Delta\lambda_{FWHM}$) (corresponding to a modulation depth of ~6 dB), the required index change of silicon can be written as $\Delta n_{Si} = n_g (\delta\lambda_r/\lambda_r) / \Gamma = n_g (\lambda_{FWHM}/\lambda_r) / \Gamma = n_g /(Q\Gamma) \approx 6.5 \times 10^{-4}$, where $n_g \approx 4.0$ is the group index around 1522 nm, calculated from the measured free spectral range ($\lambda_{FSR}$) and the ring radius (R) as $n_g = \lambda_r^2 / (2\pi R \lambda_{FSR})$, and $\Gamma \approx 0.95$ is the mode confinement factor. Such index change corresponds to a carrier density change $\rho \approx 1.6 \times 10^{17}$ cm$^{-3}$ inside the modulator [30] and an injected charge $Q_c = \rho qV \approx 110$ fC, where $q = 1.6 \times 10^{-19}$ C is the charge per carrier and $V \approx 4.4$ μm$^3$ is the physical volume of modulator intrinsic region. The energy consumption per injection (0-1 transition) is therefore $Q_c V_{pp} = 55$ fJ at the swing voltage of 0.5 $V_{pp}$, and the averaged switching energy consumption for a pseudorandom signal is therefore approximately 14 fJ/bit (the 0-1 transition has a probability of 0.25 of all bit sequences [14]). The holding energy of the modulator is estimated from the DC power consumption at the on/off states. The driving voltage and DC current at the two states are 1.65 V, 0.225 mA and 1.15 V, 0.055 mA, respectively (see Figure 3(a) and Figure 4(a)). The averaged DC power is thus 217 μW and correspondingly the DC energy consumption is approximately 72 fJ/bit at 3 Gbps data rate. The total per bit energy consumption for the modulator is thus approximately 86 fJ, dominated by the DC holding energy. The power consumption of the detector, with an averaged output DC current of approximately 0.1 mA and a bias voltage of 1 V, is about 100 μW, which corresponds to a per bit energy consumption of 33 fJ. We use the external amplifier only to slightly improve the signal visualization on the oscilloscope and its power consumption is not included (open eye-diagram can be obtained without the amplifier). The optical link reported here thus has a total energy consumption of approximately 120 fJ/bit at the data rate of 3 Gbps, excluding the power of the laser source.

The optical link can be optimized to much higher data rate and its power consumption can also be further reduced. The data rate shown here is mostly limited by the free carrier lifetime in the silicon modulator [12,13,31]. However, using a pre-emphasis driving technique a much higher data rate of up to 18 Gbps has been demonstrated [13]. The power consumption can be further reduced by scaling down the modulator size [14,32]. Note that the speed of detectors on this sample is limited to 5 Gbps (the 90% - 10% fall time of the detector response is measured to be approximately 200 ps), while previously with similar designs we have obtained speed as high as 40 Gbps [24]. We suspect the problem is related to fabrication issues (especially the metallization process). The detector can be optimized to work with bias voltage less than 1 V to reduce its DC power. Low-power CMOS transimpedance amplifiers can also be directly integrated after the detector with large feedback resistance and projected power dissipation of few hundred microwatts and per bit energy consumption on the order of 20 fJ/bit [33,34].

The basic interconnect design with compact devices that we demonstrated here can be configured for a large-scale wavelength-division multiplexing (WDM) network with extremely high aggregate bandwidth [35]. On the transmitter side, the ring-resonator based modulator scheme allows highly scalable cascading of multiple modulators, each working at a slightly different wavelength. An example of such scheme has been previously demonstrated [36]. On the receiver side, similarly cascaded passive add-drop ring-resonators can be used for demultiplexing the data. A WDM receiving network of integrated germanium detectors and silicon demultiplexers has also been demonstrated very recently [24]. Combining these previously demonstrated functionalities with the integrated optic link presented here, one could in principle develop a photonic network supporting terabit/second data transfer for on-chip interconnects.

**Acknowledgement**

The authors would like to John R. Lowell from Defense Advanced Research Projects Agency (DARPA) for partially supporting this work under the DARPA Optical Arbitrary Waveform Generation Program. This work was also partially funded by the Interconnect Focus Center Research Program, supported in part by the Microelectronics Advanced Research Corporation (MARCO), and its participating companies. This work was performed in part at the Cornell Nano-Scale Science & Technology Facility (a member of the National Nanofabrication Users Network) which is supported by National Science Foundation, its users, Cornell University and Industrial Affiliates. Part of the characterization was performed in Bell Laboratories, Alcatel-Lucent, and L. Chen acknowledges Christopher Doerr and Jeffrey Sinsky for helpful discussions.